\documentclass[sigconf]{acmart}

\usepackage{graphicx}
\graphicspath{{figures/fa22}}
\usepackage{multirow}
\usepackage{balance}
\usepackage{siunitx}
\usepackage{amsmath}
\usepackage{framed}

\AtBeginDocument{%
  }

\setcopyright{acmcopyright}
\copyrightyear{2018}
\acmYear{2018}
\acmDOI{XXXXXXX.XXXXXXX}

\acmConference[Conference acronym 'XX]{Make sure to enter the correct
  conference title from your rights confirmation emai}{June 03--05,
  2018}{Woodstock, NY}
\acmPrice{15.00}
\acmISBN{978-1-4503-XXXX-X/18/06}

\usepackage{titlesec}
\titlespacing{\section}{1pt}{2pt plus 2pt minus 2pt}{2pt plus 2pt minus 2pt}
\titlespacing{\subsection}{1pt}{2pt plus 2pt minus 2pt}{2pt plus 2pt minus 2pt}
\titlespacing{\subsubsection}{1pt}{2pt plus 2pt minus 2pt}{2pt plus 2pt minus 2pt}
\newcommand{\pp}{Parsons problems}
\newcommand{\tool}{Proof Blocks}

\newcommand{\pl}{OHEP}
\newcommand{\groupOT}{the Off-topic group}
\newcommand{\GroupOT}{The Off-topic group}
\newcommand{\groupND}{the No Distractors group}

\newcommand{\groupD}{the Distractors group}

\newcommand{\OT}{Off-topic}
\newcommand{\ND}{No Distractors}
\newcommand{\D}{Distractors}

\begin{document}

\title
[Learning Gains of Proof Blocks with Distractors]
{Measuring the Impact of Distractors on Student Learning Gains while Using Proof
Blocks}

%\author{Seth Poulsen, Yael Gertner, Benjamin Cosman, Matthew West, Geoffrey L.
%Herman}
%\email{{sethp3, ygertner, bcosman, mwest, glherman}@illinois.edu}
%%\orcid{1234-5678-9012}
%\affiliation{%
%	\institution{University of Illinois Urbana-Champaign}
%	%  \streetaddress{P.O. Box 1212}
%	%  \city{Dublin}
%	%  \state{Ohio}
%	%  \postcode{43017-6221}
%	\country{}
%}
%
\author{Seth Poulsen}
\email{sethp3@illinois.edu}
%\email{ygertner@illinois.edu}
\affiliation{%
	\institution{University of Illinois Urbana-Champaign}
	%  \streetaddress{P.O. Box 1212}
	\country{}
}

\author{Hongxuan Chen}
\email{hc10@illinois.edu}
\affiliation{%
	\institution{University of Illinois Urbana-Champaign}
	%  \streetaddress{P.O. Box 1212}
	\country{}
}

\author{Yael Gertner}
\email{ygertner@illinois.edu}
\affiliation{%
	\institution{University of Illinois Urbana-Champaign}
	%  \streetaddress{P.O. Box 1212}
	\country{}
}

\author{Benjamin Cosman}
\email{bcosman@illinois.edu}
\affiliation{%
	\institution{University of Illinois Urbana-Champaign}
	%  \streetaddress{P.O. Box 1212}
	\country{}
}

\author{Matthew West}
\email{mwest@illinois.edu}
\affiliation{%
	\institution{University of Illinois Urbana-Champaign}
	%  \streetaddress{P.O. Box 1212}
	\country{}
}

\author{Geoffrey L. Herman}
\email{glherman@illinois.edu}
\affiliation{%
	\institution{University of Illinois Urbana-Champaign}
	%  \streetaddress{P.O. Box 1212}
	\country{}
}
%
% \author{Anon}
% \email{anon@anon.edu}
% %\orcid{1234-5678-9012}
% \affiliation{%
% 	\institution{University of Anonymous}
% 	\country{Anon. Location}
% }
% \author{Anon}
% \email{anon@anon.edu}
% %\orcid{1234-5678-9012}
% \affiliation{%
% 	\institution{University of Anonymous}
% 	\country{Anon. Location}
% }\author{Anon}
% \email{anon@anon.edu}
% %\orcid{1234-5678-9012}
% \affiliation{%
% 	\institution{University of Anonymous}
% 	\country{Anon. Location}
% }\author{Anon}
% \email{anon@anon.edu}
% %\orcid{1234-5678-9012}
% \affiliation{%
% 	\institution{University of Anonymous}
% 	\country{Anon. Location}
% }\author{Anon}
% \email{anon@anon.edu}
% %\orcid{1234-5678-9012}
% \affiliation{%
% 	\institution{University of Anonymous}
% 	\country{Anon. Location}
% }\author{Anon}
% \email{anon@anon.edu}
% %\orcid{1234-5678-9012}
% \affiliation{%
% 	\institution{University of Anonymous}
% 	\country{Anon. Location}
% }

%\renewcommand{\shortauthors}{Poulsen et al.}
\renewcommand{\shortauthors}{Seth Poulsen, Hongxuan Chen, Yael Gertner, Benjamin
Cosman, Matthew West, \& Geoffrey L. Herman}
\renewcommand{\shortauthors}{Anon.}

\begin{abstract}
\noindent
\textbf{Background}:
Proof Blocks is a software tool that enables students to construct proofs by
assembling prewritten lines and gives them automated feedback. Prior work on
learning gains from Proof Blocks has focused on comparing learning gains from
Proof Blocks
against other learning activities such as writing proofs or reading.
\\ \noindent
\textbf{Purpose}: The study described in this paper aims to compare learning
gains from different variations of Proof Blocks. Specifically, we attempt to
quantify
the
difference in learning gains for students who complete Proof Blocks problems with
and
without distractors.
\\ \noindent
\textbf{Methods}:
We conducted a randomized controlled trial with three experimental groups:
a control group that completed an off-topic Proof Blocks activity, one that
completed a
\tool{} activity without distractors, and one that completed a Proof Blocks
activity with distractors. All three groups read a book chapter on proof by
induction before completing their activity.
\\ \noindent
\textbf{Findings}:
The group that completed the Proof Blocks activity with distractors performed
better on the posttest than the group that completed the Proof Blocks without
distractors, who in turn performed better than the group that completed the
off-topic Proof Blocks activity. However, none of these differences were
statistically significant. While the results of this study are inconclusive, we
hope that it can serve as a foundation for future work.
\end{abstract}


%\ccsdesc[500]{Social and professional topics~Computing education}
%\ccsdesc[500]{Applied computing~Computer-assisted instruction}


\maketitle

\section{Introduction and Background}
\begin{figure}\includegraphics[width=\columnwidth]{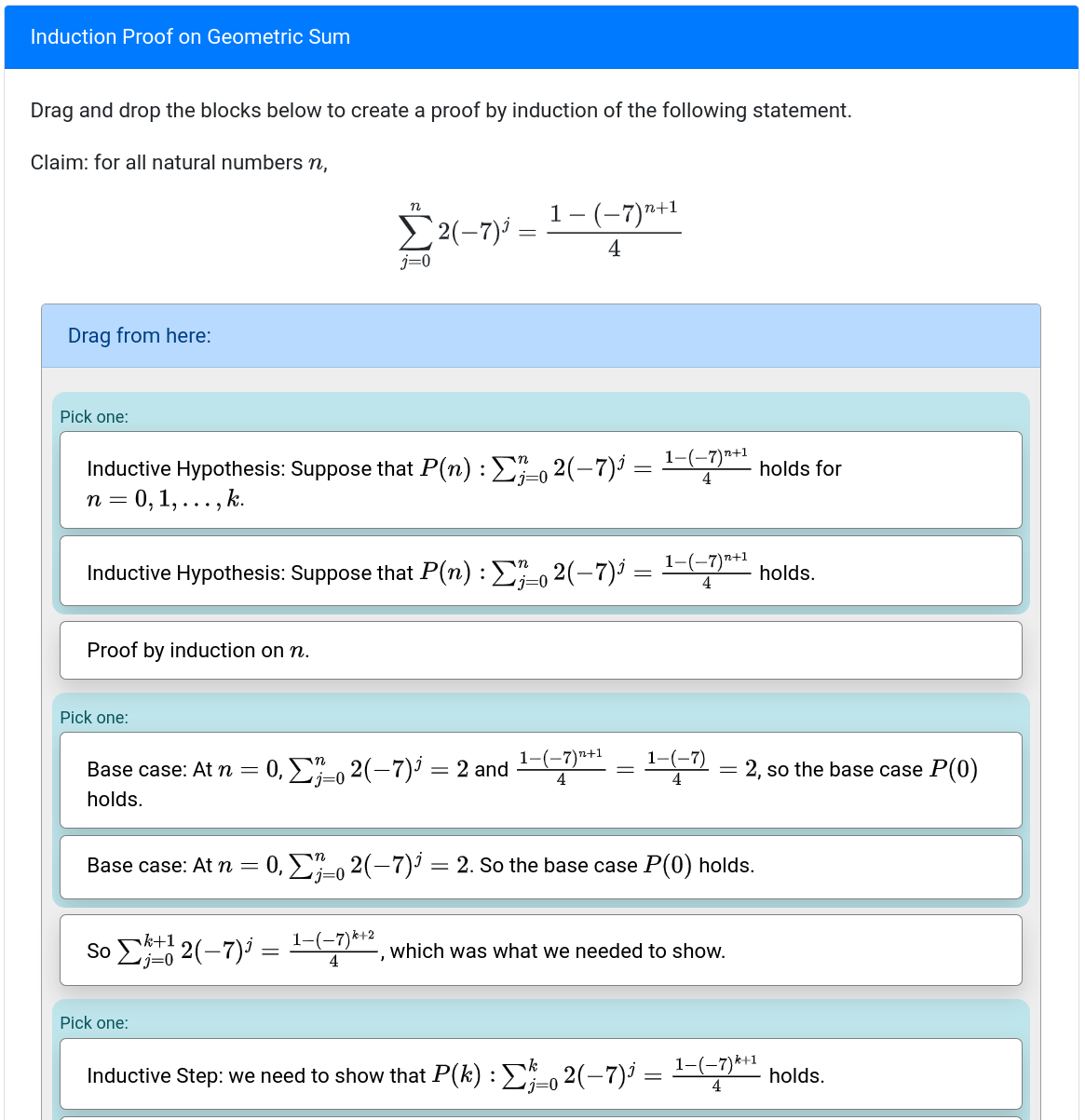}
\caption{Example \tool{} problem from the distractor learning activity.}
\label{fig:example}
\vspace{0em}
\end{figure}
Over the last few years, \tool{} have become an increasingly popular way to
support students through the process of learning to write mathematical proofs.
\tool{} scaffold the process of students learning to write proofs by allowing them
to drag and drop prewritten lines instead of needing to come up with the proof
completely on their own. Many students need support in learning to write proofs as
they struggle to write proofs even when they have the required content
knowledge~\cite{weber2001student}.

Prior work has shown that \tool{} problems used as test questions provide similar
information about student knowledge as proof writing questions and are correlated
with them~\cite{poulsen2021evaluating}. Multiple studies have also been published
that seek to measure the learning gains of Proof Blocks in comparison with other
learning activities. A randomized controlled
trial showed that students who had read a book chapter and completed \tool{}
problems learned as much as students who read a book chapter and wrote proofs from
scratch, but in less time~\cite{poulsen2023efficiency}. Another experiment showed
that students completing a reading and \tool{} performed marginally better on a
post test than students who completed the reading
alone~\cite{poulsen2024disentangling}. This paper is the first attempting to
measure the comparative learning gains of \emph{different kinds} of \tool{}
problems. Specifically, we seek to answer the following research question:
\begin{itemize}
\item[\textbf{RQ}] Do students learn more from completing \tool{} problems that
contain distractors, versus completing identical \tool{} problems that do not
contain distractors?
\end{itemize}

\section{Related Work}
There is a broad set of work related to the present work. In addition to the
extant body of literature on \tool{}, we draw inspiration from work on
mathematical proof education as well experiments designed to measure learning
gains from \pp, which are similar to \tool{} but for learning to
write code. We cover these all here. This largely overlaps the related work
mentioned in prior work on Proof Blocks, with some
additions~\cite{poulsen2023efficiency,poulsen2024disentangling}.

\subsection{Cognitive Conflict}

Cognitive conflict arises when a student holds conflicting ideas in their mind
which they must resolve, and has been studied in many educational contexts since
first appearing in Piaget's theory of development~\cite{limon2001on}. The role of
cognitive conflict has been studied in mathematics education over the last few
decades, showing varying levels of effectiveness in helping student
learning~\cite{gal2019use,behr1990students}. Most closely
related to our work,
cognitive conflict has been used to help students understand the need for
mathematical proofs~\cite{stylianides2009facilitating}. As far as we are aware, it
has not been used in the context of helping students actually learn to write
mathematical proofs.
Our theory is that by exposing students to distractor lines in their \tool{}
problems that directly relate to common student misconceptions about proof by
induction, we will help them to experience a state of cognitive conflict, which
will eventually be resolved when they either figure out or are told which version
of the proof line is correct, thus helping them properly assimilate the knowledge.
More detail on the distractors shown is given in
Section~\ref{sec:misconceptions-distractors}.

\subsection{Research on Teaching and Learning Proofs}
%\subsubsection{Interventions for learning to write proofs }
Based on their review of the literature on teaching and learning proofs,
Stylianides and Stylianides concluded that ``more intervention-oriented studies in
the area of proof are sorely needed ~\cite{stylianides2017research}.''
Most studies on how students learn about proofs have focused only on whether
students can read and comprehend
proofs~\cite{weber2012generic,malek2011effect,roy2014evaluating}. They stop short
of helping students write new proofs. For example, Hodds et
al.~\cite{hodds2014self}
showed that training students to engage more with proofs by using self-explanation
methods increased student comprehension of proofs in a lasting way. There have
also been interventions that focus on helping students understand the need for
proofs, as students often believe empirical arguments without seeing the need for
proof~\cite{stylianides2017research,jahnke2013understanding,brown2014skepticism,
stylianides2009facilitating}. There are also various experience reports
on novel interventions that instructors have tried with limited evidence on their
effectiveness~\cite{harel2001development,larsen2008proofs,norton2022addressing}.
We unfortunately have limited examples of evidence-based methods for helping
students learn proofs. A review of other
software tools with visual methods of constructing
proofs is given in prior work~\cite{poulsen2022proof}.

\subsection{Drag-and-drop and Parsons Problems}
Drag-and-drop tools are used in a variety of disciplines and tasks such as
ordering events along a timeline or ranking items according to some criteria.
\tool{} was particularly inspired by \pp~\cite{parsons2006parson}. These use
drag-and-drop tools for students to
construct programs. Both \pp{} and \tool{}
arrange blocks of text according to a logical order and enable testing of students' ability while providing some scaffolding.

\pp{} have generally been shown to be useful for both assessment and
learning~\cite{denny2008evaluating,du2020review,ericson2017solving,
ericson2018evaluating,weinman2021improving}.
Like \tool{}, \pp{} can provide rich information about student
knowledge while being easier to
grade~\cite{poulsen2021evaluating,denny2008evaluating}. Ericson et
al.~\cite{ericson2017solving,ericson2018evaluating} used randomized controlled
experiments to demonstrate that students could learn just as much from \pp{} as
they could from writing code from scratch, but in less time. \pp{} have also been
evaluated as active learning activities in
lecture~\cite{ericson2022adaptive}.

The use of distractors is common in \pp{} and generally use
distractors focused on syntax
errors~\cite{smith2023discovering,denny2008evaluating,Ericson:2015,Helminen:2013,
ihantola2011two}  or on program structure~\cite{Harms:2016}.
Adding distractors does make the \pp{} more difficult and appears to increase the
cognitive load they
induce~\cite{Harms:2016,smith2023investigating}.
While early work on \pp{} claimed that distractors were useful for
learning,~\cite{parsons2006parson} some researchers found that they have no overall
effect on learning~\cite{Harms:2016}. More recent research has shown convincing
evidence using a randomized controlled trial that distractors can improve learning
gains for students in an introductory programming class seeking to learn new
concepts~\cite{smith2024comparing}.

\section{Experimental Design}
\begin{table}
\centering
\resizebox{\columnwidth}{!}{
\begin{tabular}{c|c|c}
\OT{} $(n=63)$ & \ND{} $(n=72)$ & \D{} $(n=66)$ \\ \hline \hline
Pretest & Pretest & Pretest \\ \hline
Book Chapter & Book Chapter &	Book Chapter  \\ \hline
Off-topic & \tool{}  & \tool{} \\
\tool{} & without Distractors & with Distractors \\ \hline
Practice Test & Practice Test & Practice Test \\
\hline \hline
\multicolumn{3}{c}{One Week}  \\
\hline \hline
Posttest & Posttest & Posttest
\end{tabular}
}
\caption{Design of the learning experiment. Students were free to move on
as soon as they finished a particular portion of the activity}
\label{tab:studydesign}
\end{table}

We use a very similar methods and experimental design as those used in prior
studies of the learning gains of
\tool{}~\cite{poulsen2023efficiency,poulsen2024disentangling}, with some minor
changes.
We repeat all details here for completeness and clarity.
We used a between-subjects experimental design.
To control for confounding variables, we ran
our study as a controlled lab study rather than as part of a course.
We recruited students from the Discrete Mathematics course in our department
who already had knowledge of some proof techniques and used the textbook and/or
\tool{} to teach them a new proof technique.
All students who participated completed a pretest and read a book chapter about
proof by induction. They then completed a \tool{} learning activity, which was
either completing \tool{} problems on topics unrelated to proof by induction (we
call this group ``\groupOT{}'', abbreviated as ``\OT{}'' in tables),
completing \tool{} problems about proof by induction without distractors (we call
this group ``\groupND'',  abbreviated
as ``\ND{}'' in tables), or completing proof by induction problems that contain
distractor lines reflecting common misconceptions about proof by induction as
outlined in Section~\ref{sec:misconceptions-distractors}
(we call this group ``\groupD{}'' abbreviated as ``\D{}'' in tables).
Participants then took a practice test. This experimental design is visualized in
Table~\ref{tab:studydesign}.
As soon as a student finished the learning activity they were allowed to proceed
to the practice test.
They were given 1 hour and 50 minutes to complete their learning activity and
practice test. We also
invited them all to participate in a posttest one week later, which they were
given 50 minutes to complete. The pretest, practice test and posttest were
identical, containing
and the same two proof-writing problems.
While the students were shown the
example solutions for all exercises they completed during the learning
activities, they were not shown solutions for any of the test
questions.

\subsection{Experimental Environment}
All students completed their learning activities and tests in %\pl,
an Open source Homework and Exam Platform which we call \pl{} for
anonymization purposes
%a problem-driven online learning system
~\cite{prairielearn}.
\pl{} automatically keeps track of
when a user opens and closes each assessment and when they submit an answer to
each problem. Therefore, we were able to track and analyze the
amount of time that students spend on each portion of the study.
Beginning in Fall 2021, all sections of Discrete Mathematics have completed
homework and exams through \pl{}, so the students we recruited
were already familiar with the platform.
% and its user interface.
For the written proofs,
%in the practice test and posttest,
students wrote their proofs in a text entry box
that supported markdown and LaTeX, but were told that using plain text
%(for example, spelling out ``sum from i=0 to n" instead of using $\sum_{i=0}^n$)
was acceptable. They were
not expected to learn LaTeX for the course.
To control the student learning environment for our study, we used our
university's computer testing center, which provides a closed
environment where students can only access the
assessment that they are working on.
%~\cite{zilles2019every}
Students could choose to complete the learning activity at any
point over the period of a few days.

%To control the student learning environment for our study, we used the
%University's Computer Based Testing Facility
%(CBTF)~\cite{zilles2019every}, which provides a proctored and locked down
%environment where student work on lab computers where they can only access the
%assessment which they are supposed to be working on. The CBTF allows flexible
%scheduling so that students can choose to complete the learning activity at any
%point over the period of a few days.

\subsection{Experimental Subjects}
Students in the Discrete Mathematics course in our department learn proof by
induction in the middle of the semester. Thus, we could recruit
students for our study a few weeks into the semester, after they have
learned the basics of writing proofs, but before their course has covered
proof by induction. Since all parts of the experiment were complete before the
students covered proof by induction in class, the students had little
to no motivation to study the material outside of the context of the study,
helping with the validity of the experiment. The Discrete Mathematics course in
our department is typically taken by students early in the computer
science and computer engineering majors or computer science minor.
Introductory programming and calculus are prerequisites.

We offered students 0.5\% extra credit in the course for each day of the study
(one for the learning activity, another for the posttest). Due to
constraints of \pl, we pre-assigned all
eligible students to an experimental condition before they elected to
participate. This resulted in a small variation in the population sizes for
each treatment. We had 322
students participate in the learning activity.  As allowed by our
research protocol, 65 subjects (20.2\%) in this pool opted to not have their data
used for the research project. We threw out the data for 4 students who, by a
\pl{} configuration error, were given access to multiple of the learning activities
instead of only one. Of the 253 eligible students who consented, 201
(79.4\%) completed the posttest.
Broken down by experimental condition,
72 of 88 (81.9\%) students who started in \groupND,
63 of 81 (77.8\%) students who started in \groupOT, and
66 of 84 (78.6\%) students who started in \groupD{} were included in the
final data set.

\subsection{Test Materials}
The practice test (and the identical posttest) consisted of two written induction
proof questions. The two topics of inductive proofs tested were
(1) proving the closed form for a summation
$( \sum_{i=0}^n i = \frac{n(n+1)}{2})$
(2): proving the closed form for a
recursively-defined function (see
Figure~\ref{fig:test-2}). This is exactly the same as the test given in the study
measuring learning gains from \tool{} compared to written
proofs~\cite{poulsen2023efficiency}.

\begin{figure}
\begin{framed}
\raggedright
Suppose that $g: \mathbb{N} \rightarrow \mathbb{R}$ is defined by
\begin{align*}
g(0) &= 0 \\
g(1) &= \frac{4}{3} \\
g(n) &= \frac{4}{3}g(n-1) - \frac{1}{3}g(n-2), \text{ for } n\geq 2
\end{align*}
Use induction to prove that $g(n) = 2 - \frac{2}{3^n}$ for any natural number $n$.
\end{framed}
\caption{Question 2 from the test used in the study}
\label{fig:test-2}
\end{figure}

\subsection{Learning Activity Materials}
To check for the comparability of experimental groups, at the
beginning of the learning activity students were asked a single
question to gauge their level of familiarity with proof by induction: ``What
was your level of familiarity with proof by induction before today?'' with
answer choices (a) I was very familiar with proof by induction, (b) I was
somewhat familiar with proof by induction, and (c) I had never heard of proof
by induction.

The book chapter that students read was Sections 11.1 through 11.5 of ``Building
Blocks of Theoretical Computer
Science''~\cite{fleck2013building}, the same textbook that students were reading
for the course. Though the students may have read the first few chapters of the
book
at the point they participated in the study, the students were still a few weeks
away from covering proof by induction in class at the time of the study and so we
find it extremely unlikely that any of them had read this chapter of the book.
The book chapter talks about motivation for and
reasoning about proof by induction, as well as working through a few examples. One
of these was identical to problem 1 on the test, the other was a proof by
induction problem showing that a given algebraic expression was always divisible
by a certain integer.

The five \tool{} questions given in the learning activity consisted of three
problems similar to the first written proof and two problems similar to the second
written proof on the test.
Students were given instant feedback on their work, including
which line of their proof was the first incorrect line. This type of feedback is
commonly used in \pp{}, and it has
been called relative line-based feedback~\cite{du2020review}.
Students in \groupD{} were also given targeted feedback about specific
misconceptions if they picked a distractor line (see
Figure~\ref{fig:distractor-feedback}).
In the \pp{}
literature these have been referred to as ``paired'' or ``grouped''
distractors~\cite{ericson2017solving,smith2023comparing}.
For more details of the \tool{} autograder and feedback system,
see prior work~\cite{poulsen2022proof,poulsen2023efficient}.

Once they finished the problem, they were shown an example solution as well as
explanations about why each of the distractors was incorrect.
In all conditions, students were given three tries to complete each \tool{} before
they were shown the example solution.

\subsubsection{Distractors in Proof by Induction }
\label{sec:misconceptions-distractors}
\begin{figure*}\includegraphics[width=.8\textwidth]{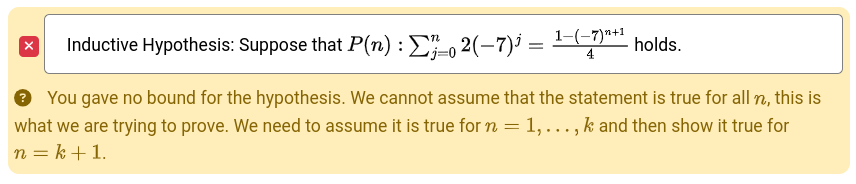}
\caption{Example feedback shown to students about a distractor line intended to
teach them about the common mistake of not bounding the inductive hypothesis.}
\label{fig:distractor-feedback}
\vspace{0em}
\end{figure*}

The distractors that we choose to show students in our \tool{} problems are based
on common misconceptions that students have about proof by induction, based on
findings from prior research as well as our own experience
%Distractors werechosen to match mistakes that we often saw students make in
%grading the
%free-response proofs from the prior studies,
%such as
These included (1) conflating the inductive
hypothesis with the proof goal or failing to bound the inductive hypothesis
(documented by Norton et al.~\cite{norton2022addressing}), (2) incorrectly
identifying base cases (documented by Baker~\cite{baker1996students}), (3)
failure to adequately prove base cases (documented by Stylianides et al. and
Baker~\cite{stylianides2007preservice,baker1996students}), and (4) failure to
understand what it means to apply the inductive hypothesis in context.
As shown in Figure~\ref{fig:example}, as students worked on \tool{} problems, they
were shown distractor lines visually paired with the corresponding correct line.
If a student selected a distractor line, they were given specific
feedback about that distractor as shown in Figure~\ref{fig:distractor-feedback}.

\section{Methods}
%\begin{table}
%\centering
%\begin{tabular}{p{2cm}|p{5.5cm}}
%Proof Section & Proof Detail \\
%\hline
%\hline
%Base Case(s)
%& (1) Identify Base Case(s) \\
%& (2) Prove Base Case(s)	\\
%\hline
%Inductive
%& (3) Hypothesis is stated	\\
%Hypothesis
%& (4) Hypothesis is given some bound	\\
%\hline
%Inductive Step
%& (5) Goal is Clear	\\
%& (6) Expression of Size $k+1$ is decomposed into expression of size
%$k$	\\
%& (7) Inductive Hypothesis is applied	\\
%\end{tabular}
%\caption{Rubric for grading written proofs. Each detail on the rubric is
%assigned the following points: 0 for not
%present, 1 for partially correct, or 2 for correct. We validated our rubric by
%having multiple authors grade the same proofs.}
%\label{tab:codebook}
%\vspace{-1em}
%\end{table}

\subsection{Rubric and Grading}
To grade the free response proof questions, we used the same 7-point proof by
induction rubric
created and validated as a part of a prior study~\cite{poulsen2023efficiency}.

At the start of grading our data, we calibrated our
research team by having all members grade the same proofs and calculating our
inter-rater reliability scores.
All four members of the grading
team (Authors 1-4 on this paper) graded ten student submissions for each of the
two proof questions used for the pilot study. We used Krippendorff's alpha to
calculate an inter-rater reliability of $0.88$
over $140$ rubric points ($2 \times 10 = 20$ proofs), well above the generally
accepted threshold of
0.8~\cite{krippendorff2004reliability}.
In the final round of grading, Author 1 graded 275 proofs,
Author 2 graded 258 proofs, Author 3 graded 149 proofs, and Author 4
graded 132 proofs. The remaining 2 proofs were automatically assigned a score of 0
due to being left blank. Each proof question was scored 0, 1, or 2 on 7
different rubric categories, allowing for 14 points possible on a single proof
question and 28 points possible across the proofs on the posttest. For ease of
reporting and understanding effect sizes, we have converted all scores to
percentages.

\section{Results}
\label{sec:results}

\subsection{Comparability of Experimental Groups}

Because students were randomly assigned to experimental conditions, we had
strong reason to believe a priori that the populations of
students in each experimental group were comparable, but we still ran
statistical checks.
A Shapiro-Wilk test showed that the pretest scores were non-normal ($p<.001$
for all three experimental groups), so we use a Kruskal-Wallis test
for this comparison. We fail to reject the null hypothesis that the
distribution of pretest scores between groups are the same ($\chi^2=1.30$,
$p=0.52$). More details of the pretest scores are shown in
Figure~\ref{fig:all-proofs} and Table~\ref{tab:preposttests}.
Next, we compared the student responses to the survey on the familiarity with
proof by induction. A Shapiro-Wilk test showed the data to be non-normal
($p<.001$ for all three experimental groups), so we used a Kruskal-Wallis test
to confirm that the familiarity level is similar across groups.
We fail to reject the null hypothesis that the distribution
of familiarity scores between groups are the same ($\chi^2=3.4$, $p=0.18$).
Details of the familiarity survey results can be seen in
Table~\ref{tab:familiarity}.
\begin{table}
\resizebox{\columnwidth}{!}{
\begin{tabular}{r|c|c|c}
Level of Familiarity& \multicolumn{3}{c}{Experimental Condition Group} \\
\cline{2-4}
with Induction &
                     \OT   &     \ND     & \D \\ \hline
Very Familiar  	  &   6    &      3      &        6                 \\
Somewhat Familiar &   17   &      22     &        26                \\
Never Heard       &   49   &      38     &        34
\end{tabular}
}
\caption{Breakdown of prior students knowledge by experimental group. Each of the
experimental groups started out roughly equal in knowledge of proof by induction
}
\label{tab:familiarity}
\end{table}

\subsection{Learning Gains}
\subsubsection{Within Group Pretest to Posttest Gain}
\begin{figure*}
\includegraphics[width=\textwidth]{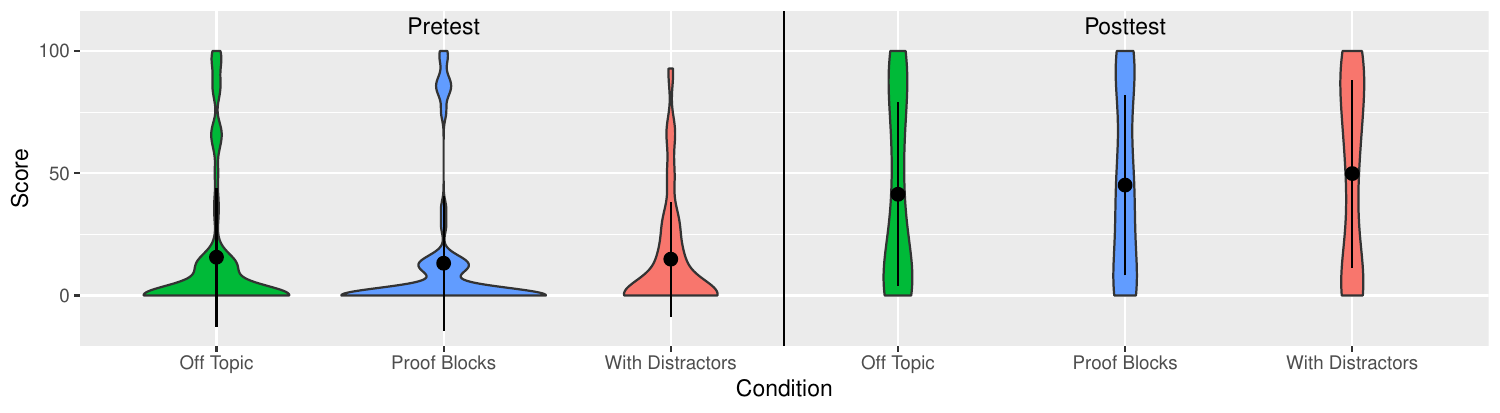}
\vspace{-1em}
\caption{Distribution of scores on all proof problems for all three
experimental groups.}
\label{fig:all-proofs}
\vskip -1em
\end{figure*}

To measure learning gains within groups, we compare the pretest
scores of each group to the posttest scores of the same group. These score
distributions can be seen in Figure~\ref{fig:all-proofs}, with summary
statistics in Table~\ref{tab:preposttests}.
\begin{table*}
\begin{tabular}{c|cccc}
& Pretest (20\%, 80\%)
& Posttest (20\%, 80\%)
& Score Increase
%& Wilcox's $Q$
& Activity Time, minutes (s.d.)
\\ \hline
\OT{} ($n=72$) & 15.7 (0, 14.3) & 41.4 (0, 85.7) & 25.7 & 7.7 (3.9) \\
\ND{} ($n=63$) & 13.2 (0, 14.3) & 45.2 (0, 90) & 32.0 & 10.5 (4.5) \\
\D{} ($n=66$) & 14.9 (0, 28.6) & 49.9 (0, 92.9) & 35.0 & 19.3 (9.3) \\
\end{tabular}
\caption{Mean, 20\% and 80\% quantiles for the pretest and posttest scores (out
of 100), score improvement between the pretest and posttest, and time spent on the
learning activity for each
experimental group.}
\label{tab:preposttests}
\vskip -1em
\end{table*}
A Shapiro-Wilk test also showed that the posttest scores were non-normal
($p<.001$ for all three experimental groups), so we use a paired Mann Whitney U
test to test for learning gains within each group and found that all three
groups performed significantly better on the posttest than on the pretest ($p <
.001$ for all three groups). Groups improved by between 25\% and 35\%. The fact
that even the group which completed \tool{} problems which were not about proof by
induction had a large score improvement from the pretest to the posttest suggests
that a significant amount of the pretest to posttest learning gains actually
happens from the reading of the book chapter. This is consistent with the
results shown in prior work comparing learning gains from reading to learning
gains from \tool{}~\cite{poulsen2024disentangling}.

\subsubsection{Between Group Learning Gains}
In order to estimate the differential learning gains across experimental groups,
we fit a regression model which uses the students' experimental group to predict
their posttest performance, controlling for prior knowledge.
We chose to use pretest score are our control for prior knowledge, because it
has a much higher correlation with posttest scores ($\tau = 0.24$) than their
familiarity survey scores ($\tau = 0.11$), and provided an overall much better
fitting model ($R^2=0.23$) than using the familiarity survey as a control did
($R^2=0.13$).
In other words, students \emph{actual}
prior knowledge predicts their performance better than their \emph{perceived}
prior knowledge. We considered using both pretest scores and the familiarity
survey as controls for prior knowledge, but with the inclusion of both resulted in
a  model that did not have any better prediction accuracy than only including
pretest scores as the control ($R^2=0.23$).

%Due to collinearity,
%we could not use both pretest score and the familiarity survey data as controls
%for prior knowledge (they are correlated at Kendall's $\tau = 0.42$). Without
%being able to use both,

The results of this regression analysis are shown in Table~\ref{tab:coefficients}.
While prior knowledge remains the strongest predictor of posttest performance,
doing \tool{} problems about proof by induction is associated with a 5.4\%
increase in posttest performance over off-topic \tool{} problems, and doing
\tool{} problems with distractors is associated with a 9\% increase in posttest
performance over off-topic \tool{} problems. However, the standard errors on these
estimates are large, and the predictions these values do not reach the standard
threshold for statistical significance.

\begin{table}
\begin{center}
\begin{tabular}{l c c c}
 & coefficient &  std. error & $p$-value \\
\hline
Intercept (\OT) & $30.99^{*}$ & $(4.16)$ & <0.001 \\
\ND{}           & $5.40$      & $(5.75)$ &  0.348   \\
\D{}            & $9.00$      & $(5.67)$ & 0.114   \\
Pretest         & $0.67^{*}$  & $(0.09)$ & <0.001 \\
\hline
R$^2$                     &&   & $0.23$        \\
Adj. R$^2$                &&   & $0.22$        \\
Num. obs.                  &&  & $201$         \\
\hline
\multicolumn{2}{l}{\scriptsize{$^{*} \text{ significant at } p<0.05$}}
\end{tabular}
\caption{Regression model predicting posttest performance based on experimental
condition, with a control for prior knowledge. The group that completed the
\tool{} activity with distractors performed better on
the posttest than the group that completed the \tool{} without distractors, who in
turn performed better than the group that completed the off-topic \tool{}
activity. However, none of these differences were statistically significant }
\label{tab:coefficients}
\end{center}
\end{table}

\subsection{Time Spent on Learning Activity}
\begin{figure}
\includegraphics[width=0.9\columnwidth]{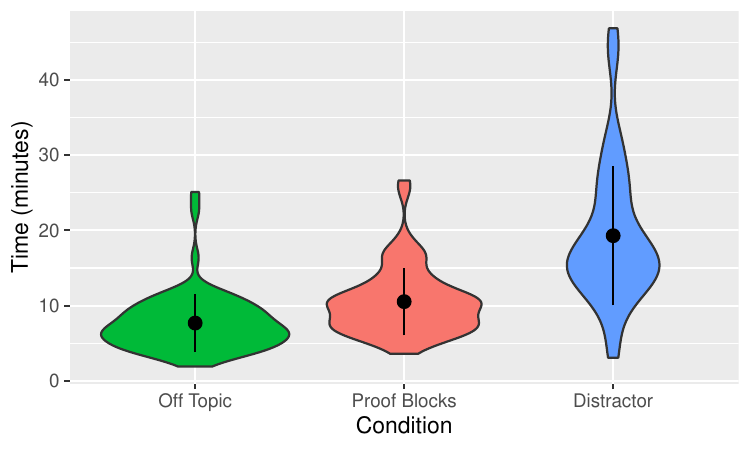}
\caption{Distribution of time spent on each learning activity by each of
the experimental groups. \GroupOT{} had a mean of 7.7 (s.d. 3.9)
minutes spent on the activity, \groupND{} spent a mean of 10.5
(s.d. 4.5) minutes, and \groupD{} spent 19.3 (s.d. 9.3) minutes. All differences
were statistically significant at $p<0.001$}
\label{fig:activity-times}
\end{figure}

\GroupOT{} had a mean of 7.7 (s.d. 3.9)
minutes spent on the activity, \groupND{} spent a mean of 10.5
(s.d. 4.5) minutes, and \groupD{} spent 19.3 (s.d. 9.3) minutes.
Pairwise $t$-tests show that the time differences between each group were all
significant at $p<0.001$.
The time distributions are shown graphically in Figure~\ref{fig:activity-times}.

%\subsection{Post-hoc Additional Analyses}
%\begin{itemize}
%\item IRT for each proof item ?
%\item beta-binomial distributions thing ?
%\end{itemize}

\section{Discussion and Limitations}

While there is clear evidence on the utility of \tool{} problems as exam questions
~\cite{poulsen2021evaluating}, this paper along with prior
work~\cite{poulsen2023efficiency,poulsen2024disentangling} the best way to use
\tool{} as a learning tool remains an open question.
Both cognitive conflict theory and recent results in
the study of learning gains from \pp{}~\cite{smith2024comparing} with distractors
would lead us to predict that \tool{} problems with distractors would help
students to achieve higher learning gains than \tool{} problems without
distractors. However, we do not find a statistically significant difference in the
learning gains of  students completing \tool{} with distractors compared to
students who completed the same \tool{} without distractors, or even compared to
students who completed an off-topic \tool{} activity.

One possible explanation is that using distractors in \tool{} does improve the
learning gains, but the effect size is small enough that it was not detectable
using the current experimental designs. Future research could tweak the
experimental design in ways that would make it easier to detect the effect.
Because of the shape of posttest score distributions, increasing the sample size
alone may not be a practical way to make the effect
measurable~\cite{poulsen2024disentangling}. The experimental design could also be
altered by giving students longer learning activities to make the differential
learning more pronounced between conditions, making the posttest longer to get a
more precise measurement of knowledge, or using a topic aside from proof by
induction that students may learn more quickly.

Another possible explanation for our results is that there actually is no benefit
for student learning to using distractors in \tool{} problems, or that the way
that our distractors were designed were not ideal for learning. This would be
consistent with prior work showing that distractors in \pp{} were not
helpful~\cite{Harms:2016}. A final
possibility is that the way the information is displayed in the user interface is
problematic, but we find this unlikely given that recent findings with \pp{} used
the exact same user interface as we used for this study~\cite{smith2024comparing}.

Consistent with the related findings of
\pp{}~\cite{smith2023investigating,smith2023comparing}, we find that \tool{}
problems with distractors do take significantly longer for students to complete.

\section{Conclusions}
This paper makes a contribution by describing the first ever experiment designed
to measure student learning gains while completing different types of \tool{}
problems. Though the results suggest that \tool{} problems with distractors may
provide greater learning gains than those without, the difference was not
statistically significant. Though this work was inconclusive, it lays a foundation
for future work on the learning gains of different kinds of \tool{} problems.

\begin{acks}
We would like to give a huge thanks to Anon and the rest
of the staff at the computer based testing facility for helping us use their
facility to run our experiment. Anon was supported by an NSF Graduate
Research Fellowship.
\end{acks}

%\begin{acks}
%We would like to give a huge thanks to Dave Mussulman and the rest
%of the staff at the computer based testing facility for helping us use their
%facility to run our experiment. Seth Poulsen was supported by an NSF Graduate
%Research Fellowship.
%\end{acks}

%\newpage
%\balance
\bibliographystyle{ACM-Reference-Format}
\bibliography{references}

\end{document}